\documentclass[aps,preprint,amsmath,amssymb,amsfonts,nofootinbib]{revtex4}
\usepackage{epsfig}
\usepackage{graphicx}
\usepackage{dcolumn}
\usepackage{bm}
\usepackage{amsthm}
\usepackage{amsmath}
\usepackage{color}

\newcommand{\beq}{\begin{equation}}
\newcommand{\eeq}{\end{equation}}
\newcommand{\bea}{\begin{eqnarray}}
\newcommand{\eea}{\end{eqnarray}}

\begin{document}

\vskip 1cm

\title{Conformal Anomalies in Hydrodynamics}
\author{Christopher Eling$^1$}
\author{Yaron Oz$^2$}
\author{Stefan Theisen$^1$}
\author{Shimon Yankielowicz$^2$}
\affiliation{$^1$ Max Planck Institute for Gravitational Physics, Albert Einstein Institute, Potsdam 14476, Germany}
\affiliation{$^2$ Raymond and Beverly Sackler School of Physics and Astronomy, Tel Aviv University, Tel Aviv 69978, Israel}
\date{\today}
\begin{abstract}
We study the effect of conformal anomalies on the hydrodynamic description
of conformal field theories in even spacetime dimensions. We consider equilibrium curved backgrounds characterized
by a time-like Killing vector and construct a local low energy effective action
that captures the conformal anomalies. Using as a special background the Rindler spacetime we
derive a formula for the anomaly effect on the hydrodynamic pressure. We find
that this anomalous effect is only due to the Euler central charge.

\end{abstract}
%
%
\maketitle
%
%
%

\section{Introduction and Summary}

Hydrodynamics is the universal effective description of any finite temperature quantum field theory on large time and length scales \cite{Landau}. In particular, on characteristic scales $L$ much greater than the correlation length  $\ell_c$ of the microscopic theory, the dynamics of the theory can be effectively described in a derivative expansion by local conservation laws of the stress-energy tensor and global symmetry currents. The conserved charges are the hydrodynamic degrees of freedom. Quantum anomalies of the microscopic quantum field theory lead to new non-dissipative transport terms in the hydrodynamic description. For instance, chiral and mixed chiral-gravitational anomalies manifest themselves by parity violating terms proportional
to the vorticity of the fluid.

In this paper we will study how trace anomalies manifest themselves in the hydrodynamic regime of a conformal field theory (CFT).
Conformal anomalies
appear in correlation functions, e.g.
two and three point functions of the stress energy tensor in two and four dimensions,
respectively (see e.g. \cite{Erdmenger:1996yc}). There are no conformal-anomalies in odd dimensions.
They can also be
seen in the one-point function of the energy-momentum tensor if the conformal field
theory is coupled to an external metric $g$ -- the source for the energy-momentum tensor --
and gauge fields $A$ -- sources for conserved currents. The general structure of the trace anomaly in $d$ spacetime dimensions is
\begin{align}
{\cal A}(g,A) \equiv \langle T^\mu_\mu\rangle_{g,A} =-(-)^{d/2}a E_d+\sum_i c_i I_i \ ,
\label{trace}
\end{align}
where $E_d$ is the Euler-density in $d$ dimensions and $I_i$ are Weyl-invariant terms,
their number depending on the dimension.

In general, the trace anomaly (\ref{trace}) is obtained from a non-local effective action which is derived by coupling the CFT to the external sources and integrating out
the CFT \cite{Deser:1999zv}.
In the hydrodynamic regime, however,
we have a systematic derivative expansion and we expect to be able to obtain the anomaly from a local effective action. Indeed we will show that working on equilibrium curved backgrounds characterized by a time-like Killing vector, there is a unique local effective action that yields the equilibrium anomalous hydrodynamics.
The independent variables that we will use are the fluid temperature $T$ and
the background curvature. When charges are present, we also have the ratio of the chemical potential and temperature, $\mu/T$, and the background gauge field strength.

The anomalous local hydrodynamic effective action takes the form
\beq
S_{\rm anom} = -\int_0^1\, d t \int
\sqrt{-\det\big(({T/T_0})^t g\big)}\, d^d x~\ln(T/T_0){\cal A}\left((T/T_0)^t g\right) \ ,
\label{anomaction}
\eeq
where $T_{0}$ is a constant.
The form of this action follows from the Wess-Zumino procedure of integrating the anomaly \cite{Wess:1971yu,Schwimmer:2010za}.
It is precisely the same anomalous local effective action one obtains by adding a dilaton $\tau$, with the replacement
$\tau \rightarrow  -\ln(T/T_0)$ \footnote{This observation is due to A. Schwimmer.}.
An immediate consequence of (\ref{anomaction}) is that the hydrodynamic effective action for the Weyl invariant parts of the anomaly (\ref{trace})
is
\begin{align}
S_{\rm Weyl} = -\sum_i c_i  \int \sqrt{-g}\,
d^4x \ln (T/T_0) I_i \ .
\label{Weylinv}
\end{align}

Consider, for instance, the four-dimensional case.
The trace anomaly reads
\begin{align}
T^\mu_\mu = -a E_4 + c W^2 \label{trace4d}  + \kappa F^2\ ,
\end{align}
where
\beq
E_4 = R^2 - 4 R_{\mu \nu} R^{\mu \nu} + R_{\mu \nu \lambda \sigma} R^{\mu \nu \lambda \sigma} \ ,
\eeq
$W^2$ is the Weyl tensor squared
\beq
W^2 = \textstyle{{1\over 3}} R^2 - 2 R_{\mu \nu} R^{\mu \nu}
+ R_{\mu \nu \lambda \sigma} R^{\mu \nu \lambda \sigma} \ ,
\eeq
and $F^2$ is the square of the gauge field strength.
When the symmetry current is the $U(1)$ R-current of an ${\cal N}=1$ supersymmetric theory,
$\kappa$ and $c$ are equal up to an overall numerical constant. In this case, $a$ and $c$ also determine the R-current anomaly.

The local effective action (\ref{anomaction}) is $S_{\rm anom}  = S_{\rm Euler}+ S_{\rm Weyl}+S_{\rm Vector}$, where
the last two actions correspond to the Weyl invariant parts of the anomaly.
One obtains
\begin{align}
S_{\rm Euler} = -a \int \sqrt{-g}\, d^4 x~
\left(\ln(T/T_0)\, E_4 -4 G^{\mu\nu}a_\mu a_\nu
+ 4\, a_\lambda a^\lambda (\nabla_\mu a^\mu)-2 \,(a_\mu a^\mu)^2 \right) \ ,
\label{effecEuler}
\end{align}
with $G^{\mu \nu} = R^{\mu \nu} -{1\over 2} g^{\mu \nu} R$ the Einstein tensor and $a_{\mu} =  -\nabla_\mu \ln T$ is the fluid
acceleration.
This yields the trace anomaly part $-a E_4$.
The Weyl invariant parts (\ref{Weylinv}) are
\begin{align}
S_{\rm Weyl} = -c \int \sqrt{-g}\,
d^4x \ln (T/T_0) W_{\mu \nu \rho \sigma} W^{\mu \nu \rho \sigma} \ ,
\label{effecWeyl}
\end{align}
which yields the trace anomaly part $c W^2$ and
\begin{align}
S_{\rm Vector} = -\kappa \int \sqrt{-g}\,
d^4x \ln (T/T_0)  F_{\mu \nu} F^{\mu \nu} \ ,
\label{effecVector}
\end{align}
which yields the trace anomaly part $\kappa F^2$.

The anomaly (\ref{trace}) arises at order $d$ in derivatives (the vector part at order $\frac{d}{2}$). However, by taking the background spacetime to be a Rindler space without a background gauge field, we will show that the anomaly has an effect also at zero order in derivatives on the hydrodynamic pressure $P$
\begin{align}
P = (\xi(\lambda_*)+ n(2\pi)^d a) T^d \label{cardyd} \ ,
\end{align}
where $\xi(\lambda)$ represents a coupling dependent term that arises from local conformally invariant terms in the effective action and $\lambda_*$ is the fixed point value of the coupling \footnote{Note that we use units where $\hbar=c=k_B=1$ throughout}. $n$ is a numerical coefficient that depends on the normalization that one chooses for the Euler
form.\footnote{Our normalization will be
$E_{2p}={1\over 2^p}R_{\mu_1\nu_1\rho_1\sigma_1}\cdots R_{\mu_p\nu_p\rho_p\sigma_p}
\epsilon^{\mu_1\nu_1\dots\mu_p\nu_p}\epsilon^{\rho_1\sigma_1\dots\rho_p\sigma_p}=R^p+\dots$.}
We will show how to calculate $n$ in arbitrary dimension and will calculate it explicitly in four and six dimensions.
Interestingly, the coefficients $c_i$ of the Weyl invariant anomalies
do not appear in the formula.

Eq. \eqref{cardyd} is the $d$-dimensional analog of the two-dimensional ``Cardy formula"
for the pressure of the CFT in the infinite volume limit \cite{Bloete:1986qm,Affleck:1986bv}, with  $\xi=0,n=1$
\begin{align}
P = 4\pi^2 c\, T^2 \ . \label{Cardy}
\end{align}
Here $c$ is the trace anomaly central charge
\begin{align}
T^\mu_\mu = c\, R \ . \label{trace2d}
\end{align}
In \cite{Jensen:2012kj} the authors argued that this jump in the derivative expansion can be explained by a ``Casimir momentum density" and computed it by considering the theory on a Euclidean cone.

The paper is organized as follows. In  section 2 we will briefly review the hydrodynamics framework, quantum anomalies and the
partition function for non-dissipative terms. We will introduce the use of Rindler spacetime as a background tool to study the effect of trace anomaly on lower order terms in the hydrodynamic derivative expansion. In section 3 we will consider the conformal anomaly in two-dimensional spacetime. We will briefly review the work of \cite{Jensen:2012kj} on effective action for $d=2$ CFT's and show explicitly how evaluating the stress-energy tensor on the Rindler metric reproduces the Cardy formula (\ref{Cardy}). In Section 4, we study the generalization to higher dimensions and derive the anomalous actions hydrodynamic actions and the formula for the pressure (\ref{cardyd}).

\section{Hydrodynamics and quantum anomalies}

\subsection{The relativistic hydrodynamics framework}

Hydrodynamics is described just by local conservation laws, with conserved charges as the low energy degrees of freedom. The most familiar example is the conservation of the energy-momentum stress tensor $T^{\mu \nu}$
\begin{align}
\nabla_\mu T^{\mu \nu} = 0 \ ,
\end{align}
where the energy density $T^{00} = \rho$ and spatial momentum $T^{0i} =\Pi^i$ are the conserved quantities. In order to have a closed system of equations, the remaining set of variables $T^{ij}$ must be determined. The hydrodynamic ansatz (constitutive relations) is to express these in terms of the energy density and spatial momentum and work order by order in an expansion in the dimensionless small quantity $\ell_{c}/L$. This amounts to
\begin{align}
T^{ij} = P \delta^{ij} + \Delta^{ij} (\partial \rho, \partial \Pi^i) \ , \label{pressure}
\end{align}
where $P$ is the pressure, which is related to the energy density via the equation of state, and $\Delta^{ij}$ contains higher order corrections depending on derivatives of the conserved charges.

One can now write the stress tensor in boost covariant form and exchange the variables $(\rho, \Pi^i)$ in favor of $(T,u^\mu)$, where $T$ is the fluid temperature and $u^\mu = (\gamma, \gamma v^i)$ is the fluid four-velocity satisfying $u_{\mu}u^{\mu}=-1$. The result is
\begin{align}
T^{\mu \nu} = \left(\rho(T)+P(T)\right) u^\mu u^\nu + P(T) g^{\mu \nu} + \Delta^{\mu \nu}(\partial T, \partial u) \ . \label{hydrostress}
\end{align}
In cases where there are additional conserved charges the hydrodynamics equations must be supplemented by an additional conservation equation
for each such charge density $n$
\begin{align}
\nabla_\mu J^\mu = 0 \ .
\end{align}
Following the same procedure as before, the current takes the form
\begin{align}
J^\mu = n u^\mu + \nu^\mu, \label{hydrocurrent}
\end{align}
where $\nu^\mu$ contains derivative corrections.

The first two terms in (\ref{hydrostress}) and the first term in (\ref{hydrocurrent}) at zeroth derivative order represent an ideal, equilibrium fluid. Indeed, using the thermodynamic relations
\begin{align}
\rho+P =&\, s\, T + \mu\, n \ ,\nonumber \\
d\rho =&\, T\, ds + \mu\, dn \ , \\
dP =&\, s\, dT + n\, d\mu \ ,\nonumber
\end{align}
where $s$ is the fluid entropy density and $\mu$ a chemical potential, one can show that the equation $u_\nu \nabla_\mu T^{\mu \nu} = 0$ can be re-expressed as the conservation of the entropy current $s^\mu$ \cite{Landau}
\begin{align}
\nabla_\mu s^\mu = \nabla_\mu (s u^\mu) = 0 \ .
\end{align}

In the standard theory \cite{Landau}, higher derivative corrections are associated with non-equilibrium physics. In this setting, the meaning of the fluid variables $(\rho, n, u^\mu)$, which were originally defined relative to the equilibrium, is ambiguous. This field redefinition ambiguity can be fixed by the choice of Landau ``frame": $\Delta^{\mu \nu} u_\nu = 0$ and $\nu^\mu u_\mu = 0$, which means higher order derivative corrections do not change the energy density and conserved charge $n$. Imposing this condition and requiring that higher order corrections to the entropy are such that the Second Law $\nabla_\mu s^\mu \geq 0$ holds, one finds
\begin{align}
\Delta^{\mu \nu} &= -2\eta\, \sigma^{\mu \nu}-\zeta P^{\mu \nu} (\nabla_\lambda u^\lambda)  \ .\\
\nu^\mu &= -\sigma_c\, T\, P^{\mu \nu} \partial_\nu \left(\frac{\mu}{T}\right) \ , \label{firstorder}
\end{align}
where $P^{\mu\nu} = g^{\mu \nu} + u^\mu u^\nu$ and  $\sigma^{\mu \nu} = P^{\mu \lambda} P^{\nu \sigma} \left(\nabla_{(\lambda} u_{\sigma)} - \frac{1}{d-1} P_{\lambda \sigma} (\nabla_\lambda u^\lambda) \right)$ is the trace-free shear tensor. We define
$\nabla_{(\mu}u_{\nu)}={1\over2}(\nabla_\mu u_\nu+\nabla_\nu u_\mu)$
and likewise (with minus sign) for $\nabla_{[\mu}u_{\nu]}$.
The transport coefficients $\eta$, $\zeta$ and $\sigma_c$ are
the shear viscosity, bulk viscosity and conductivity, respectively. To first order, the dissipative correction to the entropy current has the form
\begin{align}
s^\mu = s u^\mu - \frac{\mu}{T} \nu^\mu \label{firstent} \ .
\end{align}

Interest in hydrodynamics as an effective field theory has been re-kindled over the past several years largely due to the holographic AdS/CFT correspondence \cite{Aharony:1999ti} which states that certain conformal field theories are equivalent to (quantum) gravity on asymptotically anti-de-Sitter (AdS) space-times in one higher dimension. An interesting consequence of this duality is that the hydrodynamic regime of a strongly coupled conformal field theory (e.g. the $N=4$ super Yang-Mills theory at large $N_c$) is dual to a perturbed classical black brane solution in five-dimensional asymptotically AdS spacetime \cite{Bhattacharyya:2008jc}.  The hydrodynamics of this theory with a R-charge was studied in \cite{Erdmenger:2008rm,Banerjee:2008th} using the dual gravitational solution with a bulk Chern-Simons term. It turns out that the charge current in (\ref{firstorder}) receives an unexpected additional parity violating term $\xi \omega^\mu$ proportional to the fluid vorticity,
\begin{align}
\omega^\mu = \frac{1}{2} \epsilon^{\mu \nu \lambda \sigma} u_\nu \nabla_\lambda u_\sigma \ .
\end{align}
The existence of this term at the hydrodynamical level is a consequence of the chiral anomaly of the microscopic quantum field theory and is independent of any holographic duality
\cite{Son:2009tf,Neiman:2010zi}.  Similarly, the mixed chiral gravitational anomaly  also leads to new, non-trivial transport coefficients at hydrodynamical level \cite{Landsteiner:2011cp,Landsteiner:2011iq,Chapman:2012my}.
Note, that these various anomalous terms and new transport coefficients are \textit{non-dissipative} even though they are higher order in the derivative expansion. Positivity of the entropy current divergence provides means of constraining the transport coefficients.
In particular, it fixes uniquely the form of the chiral anomalous transport in hydrodynamics.

\subsection{Partition function}

It has recently been shown \cite{Jensen:2012jh,Banerjee:2012iz} that the relations between the non-dissipative coefficients appear because equilibrium hydrodynamics can be alternatively described by a single partition function or effective action. The constraints on the form of the stress tensor and currents arising via the variation of a Lagrangian are the same as if one introduced a conserved entropy current at the level of the equations of motion. The idea is that in hydrodynamic equilibrium, correlation functions can be obtained by the variation of a local action
\begin{align}
S = \int d^d x L(x) \ .
\end{align}
In order to be gauge and diffeomorphism invariant, the Lagrangian is made up of scalars constructed out of the background metric $g_{\mu \nu}$ and background gauge fields $A_\mu$, the temperature, any chemical potentials, and the fluid four-velocity.
One computes the stress-energy tensor and the charge current via the effective action using
\begin{align}
T^{\mu \nu} = \frac{2}{\sqrt{-g}} \frac{\delta S}{\delta g_{\mu \nu}},~~~~~J^\mu = \frac{1}{\sqrt{-g}} \frac{\delta S}{\delta A_\mu} \label{stressformula} \ .
\end{align}

The notion of equilibrium is characterized by the existence of a timelike Killing vector $V^\mu$ defined such that
\begin{align}
\mathcal{L}_V g_{\mu \nu} =& 0 \ , \\
\mathcal{L}_V A_\mu =&  0 \ ,
\end{align}
where $\mathcal{L}_V$ is the Lie derivative.
Identifying $u^\mu = V^\mu/\sqrt{-V^2}$, one can show these equations imply the vanishing of the fluid shear and expansion
\begin{align}
\sigma_{\mu \nu} = 0 \ , \\
\nabla_\mu u^\mu =  0 \ ,
\end{align}
and thus, using also $T = T_0/\sqrt{-V^2}$ with constant $T_0$ and $\mu={T\over T_0}A_\mu V^\mu$,
we have the relations
\begin{align}
\nabla_\mu u_\nu = - u_\mu a_\nu + \Omega_{\mu \nu}, \quad \nabla_\mu \ln T = -a_\mu, \quad \nabla_\mu \mu = -\mu a_\mu + E_\mu \ , \label{eqcond}
\end{align}
where the acceleration and vorticity tensor are
\begin{align}
a_\mu = u^\nu \nabla_\nu u_\mu, \quad \Omega^{\mu \nu} = P^{\mu \lambda} P^{\nu \sigma} \nabla_{[\lambda} u_{\sigma]} \ ,
\end{align}
and $E^{\mu} = F^{\mu\nu}u_{\nu}$.

Thus, we construct the effective action in the following way. In addition to the temperature and the chemical potential, we build scalars  from $a_\mu$, $\Omega_{\mu \nu}$, $E_\mu$, and invariants of the background Riemann tensor $R_{\mu \nu \lambda \sigma}$ and their derivatives. An equivalent and sometimes useful representation is to choose $V^\mu\partial_\mu = \partial_t$. One can then write the equilibrium metric in the Kaluza-Klein (KK) form (dimensionally reducing over the time coordinate) and splitting the gauge potential into time and space components \cite{Banerjee:2012iz}:
\begin{align}
ds^2 =& -e^{2 f(x)}(dt + b_i(x) dx^i)^2 + g_{ij}(x) dx^i dx^j \ , \label{KKmetric}\\
A^\mu =& (A^0, A^i) \ .
\end{align}
In this representation, $T= T_0 e^{-f}$, where $T_0$ is a constant. In addition, $a_i = \nabla_i f$ and the KK photon field $b_i$ is related to the vorticity tensor, while the full $d$-dimensional curvature is replaced by the $(d-1)$-dimensional curvature constructed from the spatial metric
$g_{ij}$.

\subsection{Conformal anomaly and Rindler space}

Consider the conformal anomaly in $d=2$ (\ref{trace2d}).
The anomaly arises here at second order in derivatives (due to the Ricci scalar), but it was shown in \cite{Bloete:1986qm,Affleck:1986bv} that there is a ``Cardy formula" for the pressure of the CFT in the infinite volume limit (\ref{Cardy}).
Thus a zeroth order, equilibrium quantity is related to the anomaly coefficient, which one would naively expect to first appear at second order hydrodynamic expansion.
In \cite{Jensen:2012kj} the authors argued that this jump in the derivative expansion can be explained by a ``Casimir momentum density" and computed by considering the theory on a Euclidean cone
\begin{align}
ds^2 = dr^2 + r^2 d\tau^2 \ ,
\end{align}
where $\tau \sim \tau+2\pi \delta$, such that $2\pi(1-\delta)$ is the deficit angle of the cone. The conical geometry (periodicity in time coordinate) induces a temperature, which depends on the value of the deficit angle and the radial coordinate
\begin{align}
T^{-1} = 2\pi \delta\, r \ \label{tempgeom}.
\end{align}
Hydrodynamics is valid as an effective theory when $|\nabla T|/T^2 \ll 1$.  Using (\ref{tempgeom}), we find the following condition on the deficit angle
\begin{align}
\delta \ll 1 \label{deficit}.
\end{align}
When computing the stress-energy tensor (\ref{stressformula}) on the conical geometry in
the limit $\delta \rightarrow 1$, all terms in the derivative expansion of the stress tensor
are of the same order and in principle non-hydrodynamic modes could contribute.
In \cite{Jensen:2012kj} it was argued that the higher derivative corrections and terms non-analytic in derivatives do not contribute to the full stress tensor in a 2d conformal theory. The requirement that both the temperature $T$ and the stress tensor vanish in this limit then lead to the Cardy formula (\ref{Cardy}). In higher dimensions, higher derivative and non-analytic terms are present and can contribute in the limit as $\delta \rightarrow 1$. We will not discuss the non-analytic terms, as they are not included in the hydrodynamic description of the system.

Considering the theory on a conical geometry also introduces a technical issue associated with the singularity at the tip. In particular, for fields with spin greater than 1, it has been explicitly found that the heat kernel (and thus the effective action) is not continuous as $\delta \rightarrow 1$ \cite{Fursaev:1996uz}. This is because higher spin fields are sensitive to the geometry and since the conical geometry breaks the symmetries of the plane, the number of zero modes is discontinuous. This suggests the possibility that relations between transport coefficients and anomaly coefficients should be dealt with care and may not hold in general. However, here in the case of the conformal anomaly we are dealing with fields of spin $\leq 1$ so this ambiguity will not play a role.
 
In this paper we will rephrase the argument of \cite{Jensen:2012kj} in a Lorentzian signature. We consider a field theory at finite temperature $T_0$ quantized on the Rindler wedge given by the metric
\begin{align}
ds^2 = -x_1^2 dt^2 + dx_i dx^i \ , \label{Rindler}
\end{align}
where $i=1,\dots,(d-1)$. In writing this metric we take the coordinate $t$ to be a dimensionless time, or (hyperbolic) angle associated with boosts. Here $T_0$ is the intrinsic dimensionless ``temperature" of the theory and not induced by the geometry itself \footnote{Note that one can go back to the standard dimensions for time and temperature by introducing a constant $\kappa$ with units of inverse length in the metric: $ds^2 = -\kappa^2 x_1^2 dt^2 + dx_i dx^i$. However, $\kappa$ is arbitrary and doesn't affect the physics.}. In this setting, the effective (Tolman) temperature is
\beq
T = \frac{T_0}{x_1} \ .
\eeq
Despite the fact that the temperature is no longer a constant, the theory is in equilibrium. In fact the form of the temperature follows from the equilibrium condition $a_\mu = -\nabla_\mu \ln T$ derived above, where $a_\mu$ is the non-zero acceleration of $V^\mu/\sqrt{-V^2}$,
where $V^\mu\partial_\mu = \partial_t$ is the timelike Killing vector of the Rindler metric. Following the same steps as above in the conical case, we find that the derivative expansion is valid when $T_0 \gg 1$.

Now the key step is to tune the temperature to the special value $T_0 = 1/2\pi$, which is Lorentzian analog of $\delta \rightarrow 1$. This is crucial because it has been proven that for any interacting field theory \cite{Unruh:1983ac,Israel:1976ur}
\begin{align}
Z^{-1} \rm{Tr} (e^{-2\pi H_R} \mathcal{O}) = \langle0| \mathcal{O} |0\rangle \ ,
\end{align}
where $H_R$ is the Rindler Hamiltonian generating boost time translations on the Rindler metric and $|0\rangle$ is the zero temperature Minkowski vacuum.
Thus, at this special temperature, the thermal expectation value of an operator $\mathcal{O}$ in a given theory on the Rindler wedge is equivalent to the zero temperature Minkowski vacuum expectation value. Taking $\mathcal{O} = T^{\mu \nu}$ and
\begin{align}
\langle0| T^{\mu \nu} |0\rangle = 0 \ ,
\end{align}
reproduces the essential features of the cone argument but without having to deal with a singular point as the Rindler metric is flat everywhere \footnote{Note on the other hand that the renormalized stress tensor associated with fields at $T_0 \neq 1/2\pi$ in the Rindler wedge is singular at the horizon}. Whether a calculation in Lorentzian signature can ameliorate the technical issue of discontinuity in the effective action for fields of spins greater than 1 remains an open question. More physically, although the Rindler metric is flat, there is an effectively constant background gravitational field which illuminates the anomaly structure.

In section 4, we will use this procedure to analyze conformal hydrodynamics with anomaly in four and higher dimensions and
derive the relation (\ref{cardyd}). Note that recently the conformal anomaly has been discussed in the context of heavy ion collisions \cite{Basar:2012bp}.

\section{Conformal hydrodynamics and anomaly in $d=2$}

In this section we consider the conformal anomaly in $d=2$ hydrodynamics. The discussion parallels the one in \cite{Jensen:2012kj} and
is a preparation for the higher dimensional analysis in the next section.
As discussed, the effective action is built from the background metric and other fields characterizing the equilibrium state. In addition to being diffeomorphism and gauge invariant, a non-anomalous effective action must also be conformally (or Weyl) invariant.
To consider the conformal anomaly, we introduce a conformally non-invariant, second order term into the effective action such that
\begin{align}
\delta_\sigma S_{\rm anom}=\int \sqrt{-g}\,\sigma\,T^\mu_\mu \ ,
\end{align}
where $\delta_{\sigma}$ is a Weyl variation of the background fields.

Under a Weyl transformation, in general dimension,
\begin{align}
g_{\mu\nu}\,\to\,\tilde{g}_{\mu \nu} = e^{2\sigma} g_{\mu \nu} \ ,
\end{align}
from which the (infinitesimal) transformation of our fields follow:
\begin{align}
\delta_\sigma g_{\mu \nu} &= 2\, \sigma\, g_{\mu \nu}\ ,\nonumber \\
\delta_\sigma u^\mu &= -\sigma\, u^\mu  \ , \nonumber \\
\delta_\sigma T &= -\sigma\, T \ , \nonumber\\
\delta_\sigma \mu &= -\sigma\, \mu \ , \nonumber \\
\delta_\sigma a_\mu &= \nabla_\mu \sigma \ , \\
\delta_\sigma \Omega_{\mu \nu} &= \sigma\, \Omega_{\mu \nu} \ , \nonumber \\
\delta_\sigma F_{\mu \nu} &= 0 \ , \nonumber \\
\delta_\sigma E_\mu &= -\sigma\, E_\mu \ .\nonumber
\end{align}

In two dimensions there are two important simplifications. The vorticity tensor $\Omega_{\mu \nu}$ is identically zero and the background Riemann curvature for the equilibrium metric (e.g. (\ref{KKmetric})) is not independent
\begin{align}
R_{\mu \nu \rho \sigma} = -\nabla_\lambda a^\lambda \left(g_{\mu \rho} g_{\nu \sigma}
- g_{\mu \sigma} g_{\nu \rho}\right) \ .\label{Riemann2d}
\end{align}
Thus, we have $T$, $\mu$, $a_\mu$, $E_\mu$ and their derivatives from which we can construct scalars. One can always trade derivatives of the temperature and the chemical potential for $a_\mu$ and $E_\mu$ using the equilibrium relations.

Up to second order in derivatives, the conformally invariant effective action in two dimensions takes the generic form
\begin{align}
S_{\rm inv}= \int d^2 x \sqrt{-g} \left(T^2\, p_0
+ {1\over T^2}\alpha E_\mu E^\mu \right) \ ,
\end{align}
where $p_0$ and $\alpha$ are functions of the Weyl invariant ratio $\mu/T$.
In the case where the theory has no charge,
\begin{align}
S_{\rm inv} = \int d^2 x \sqrt{-g}\, T^2 p_0
\end{align}
with $p_0={\rm const}$. $S_{\rm inv}$
is exact to all orders in the derivative expansion. The reason is that the Weyl tensor
$W_{\mu \nu \rho \sigma}$, which transforms homogeneously under a Weyl transformation, is
identically zero in 2d and there are no invariant scalars one can construct only from derivatives of $a_\mu$.

Consider next the conformal anomaly (\ref{trace2d}).
In general the trace anomaly is generated by a \textit{non-local} effective action. In two dimensions the conformal variation $\delta_\sigma$ of the Polyakov action
\begin{align}
S_{\rm P} \sim \int d^2 x \sqrt{-g}\, R\, \Box^{-1} R \ ,
\end{align}
yields (\ref{trace2d}) in general. However, on the equilibrium background metric (\ref{KKmetric}), $R = -2 \Box f$. This indicates that the Polyakov action may take a local form on these backgrounds. Note, however, that $\Box^{-1} \Box f = f$ only when $f$ is zero at infinity.

One procedure is to find all the possible scalars constructed from $a_\mu$ at second order in derivatives and to find the linear combination whose Weyl variation gives the anomaly.
In 2d the task is simple: the only possibility is
\begin{align}
S_{\rm anom} = c \int d^2 x \sqrt{-g}\, a_\mu a^\mu \ ,\label{Sanom2da}
\end{align}
since $\nabla_\mu a^\mu$ is a total derivative. Indeed,
$\delta_\sigma S_{\rm anom} = -2 c \int d^2 x \sqrt{-g}\,\sigma \nabla_\mu a^\mu$ which is equivalent to $c  R$ on the equilibrium background metric. Thus, the trace anomaly can be generated from a local effective action in the hydrodynamic regime.

An alternative way of writing the effective action is to use (\ref{anomaction})
\begin{align}
S_{\rm anom}=c \int d^2 x\sqrt{-g}\left(-\ln(T/T_0)R-a_\mu a^\mu\right) \ .
\label{Sanom2db}
\end{align}
The equivalence holds on the equilibrium metric backgrounds, using \eqref{Riemann2d}
and \eqref{eqcond} and integrating by parts.

The resulting equilibrium hydrodynamic stress-energy  tensor follows from the metric variation of $S=S_{\rm inv}+S_{\rm anom}$ via (\ref{stressformula}).
Note, that in addition to the explicit metrics, $a_\mu$ and $T$ also implicitly depend on the metric. One can write the stress-energy tensor up to second order in perfect fluid form
\begin{align}
T^{\mu \nu} = \rho\, u^\mu u^\nu + P\, P^{\mu \nu} \ ,
\end{align}
where
\begin{align}
\rho = p_0 T^2 - c  a_\mu a^\mu - \frac{\alpha}{T^2} E_\mu E^\mu
+ 2 c  \nabla_\mu a^\mu \\
P = p_0 T^2 - c  a_\mu a^\mu - \frac{\alpha}{T^2} E_\mu E^\mu \ .
\end{align}
Again, in the uncharged theory ($\mu=0$), the results for the energy density and pressure are exact to all orders. This result shows that, as expected, the anomaly coefficient affects the hydrodynamics at second order in the derivative expansion on a generic curved background.

We now evaluate this stress-energy tensor on the Rindler metric (\ref{Rindler}). In this case $R \sim \nabla_\mu a^\mu = 0$, but
\begin{align}
a_\mu a^\mu = \frac{1}{x_1^2} = \frac{T^2}{T_0^2} \ ,
\end{align}
has a non-trivial value. One finds
\begin{align}
\rho = P = \left(p_0 - \frac{c}{T_0^2} \right) T^2 \ .
\end{align}
So while the trace vanishes on the flat background, the anomaly still appears in the values of the energy density and pressure. Setting $T_0 = 1/2\pi$ and requiring that the stress-energy tensor vanishes yields (\ref{Cardy}) for Minkowski space,
exactly as found in \cite{Jensen:2012kj}, by putting the theory on the Euclidean cone. As before, this result is exact in the uncharged case, if a charge current is present on the Rindler background there are $O(\mu)$ corrections.

\section{Conformal hydrodynamics and anomaly in higher dimensions}

\subsection{Four dimensions}

We will first consider the generalization of the effective action approach to the case of four-dimensional CFT's. As before, the first step is to find all the conformally invariant scalars that can contribute, at a given order in derivatives.
In this case we have a proliferation of possible terms because the curvature is now an independent variable and the vorticity tensor is non-zero. At zeroth order we have just
\begin{align}
L^{(0)}_{inv} = \sqrt{-g} p_0 T^4 \ , \label{4thzeroth}
\end{align}
where $p_0$ is a function of the Weyl invariant ratio $\mu/T$.
At higher orders, one can construct conformal invariants from $F_{\mu \nu}$, which is invariant, and the vorticity, which transforms homogeneously.  Another possible tensor which transforms homogeneously is $\nabla_\mu a_\nu-\nabla_\nu a_\mu$, but for $a_\mu=-\nabla_\mu\ln T$ it vanishes.
Finally, one can also build invariants from the Riemann tensor
$\mathcal{R}_{\mu \nu \lambda \sigma}$ constructed from a Weyl covariant connection \cite{Loganayagam:2008is},
\begin{align}
\mathcal{R}_{\mu \nu \lambda \sigma} = R_{\mu \nu \lambda \sigma}
- 4\,\delta^\alpha_{[\mu} g_{\nu][\lambda} \delta^\beta_{\sigma]} (\nabla_{(\alpha} a_{\beta)}
+ a_\alpha a_\beta  - \frac{1}{2}g_{\alpha \beta} a_\lambda a^\lambda) \ .
\end{align}
It is constructed such that it is, as the Weyl tensor, invariant under
Weyl transformations
\begin{align}
\delta_\sigma \mathcal{R}_{\mu \nu \lambda}{}^{\tau}  = 0 \ .
\end{align}
At second order in derivatives in $L_{inv}$ there are the terms
$\sqrt{-g}F_{\mu\nu}F^{\mu\nu}$, $\sqrt{-g} T^2 {\cal R}$, $\sqrt{-g} T^2 \Omega_{\mu \nu} \Omega^{\mu \nu}$, and $\sqrt{-g} T F_{\mu \nu} \Omega^{\mu \nu}$. At fourth order we have e.g.
$\sqrt{-g} {\cal R}_{\mu\nu\rho\sigma}{\cal R}^{\mu\nu\rho\sigma}$, $\sqrt{-g} {\cal R}_{\mu\nu}{\cal R}^{\mu\nu}$, $\sqrt{-g} {\cal R}^2$, $\sqrt{-g} (\Omega_{\mu \nu} \Omega^{\mu \nu})^2$,
etc. All of these terms are multiplied, in general, by functions of the invariant quantity $\mu/T$, which depend on the microscopic theory.

We want to derive the trace anomaly (\ref{trace4d}) from a local effective action.
Note, that $\delta_\sigma\ln(T/T_0)=-\sigma$
and recall from \eqref{eqcond} $a_\mu=-\nabla_\mu\ln(T/T_0)$. The Weyl transformation
of $-\ln(T/T_0)$ is the same as that of the dilation $\tau$ in an spontaneously broken
CFT, for which the effective action with the dilaton can be constructed by
integrating the anomaly. The effective action in the hydrodynamic regime
is then simply obtained by the replacement $\tau\to-\ln(T/T_0)$. This yields the anomalous
actions \eqref{effecEuler}, \eqref{effecWeyl} and \eqref{effecVector}.
Note, however, that $\ln(T/T_0)$ is not an independent field from the background and
it transforms under generic variations of the metric which contribute to the
energy-momentum tensor.

There is an alternative form for $S_{\rm Euler}$, just as in 2d there were two equivalent
actions, \eqref{Sanom2da} and \eqref{Sanom2db}. This is because the Euler density $E_d$ is a total derivative in $d$ dimensions and one can integrate the $-\ln(T/T_0) E_d$ term by parts. In the case of vanishing vorticity, i.e. $b_i(x)=0$ in \eqref{KKmetric} on equilibrium backgrounds, $E_4=\nabla_\mu V^\mu$ with
\begin{align}
V^\mu = 8 (a_\lambda a^\lambda) a^\mu + 8 G^{\mu \nu} a_\nu - 8 (\nabla_\lambda a^\lambda) a^\mu + 4 \nabla^\mu (a_\lambda a^\lambda) \ .
\end{align}
Inserting this into \eqref{effecEuler} and integrating by parts leads to
\begin{align}
S_{\rm Euler}= -2 a\int\sqrt{-g}\,d^4 x\,
\left(4 a_\lambda a^\lambda (\nabla_\mu a^\mu)-2 G^{\mu\nu}a_\mu a_\nu
-3(a_\mu a^\mu)^2 \right)\,.\label{effecEuler2}
\end{align}
It is straightforward to show that this action reproduces the Euler anomaly when
evaluated on an equilibrium background with zero vorticity.

From the metric variation of $S_{\rm Euler}+S_{\rm Weyl}+S_{\rm Vector}$ one can find the contributions to the stress-energy tensor, $T_{\rm Euler}^{\mu\nu}+T_{\rm Weyl}^{\mu\nu} + T_{\rm Vector}^{\mu\nu}$,
due to the anomaly (which is at fourth order).
As $S_{\rm Weyl}$ and $S_{\rm Vector}$ contain only terms quadratic
in the curvatures of the background fields, it follows immediately that
they do not contribute in a flat background. This leaves the contribution
from $S_{\rm Euler}$. It is straightforward to evaluate
$T^{\mu\nu}_{\rm Euler}$, but the resulting expression is not very illuminating.
Its evaluation on a Rindler background is, however, very simple and we find it takes the perfect fluid form, with
\begin{align}
P_{\rm Euler} = {1\over 3}\rho_{\rm Euler} = -\frac{2a}{T_0^4} T^4 \ .
\end{align}
The zeroth order contribution from the invariant part (\ref{4thzeroth}), evaluated on Rindler space,
also yields
\begin{align}
T^{\mu \nu} = p_0 T^4 (4 u^\mu u^\nu + g^{\mu \nu})
\end{align}
so that $P_0 = p_0 T^4$ and $\rho_0=3P_0$, as expected.
Finally, the higher derivative invariant terms in the effective action also contribute,
but in a model dependent way. Since the derivative expansion breaks down in Rindler space, there are in principle an infinite number
of these terms at arbitrary order in derivatives contributing to the equilibrium pressure.
Parametrizing their contribution by a function $\xi(\lambda,T_0)$, where
$\lambda$ is a coupling constant in the microscopic theory, we obtain
\begin{align}
P_{\rm tot} = \left(p_0 - \xi(\lambda,T_0) - \frac{2a}{T_0^4} \right) T^4 \ .
\end{align}
Demanding that the stress tensor vanishes in the Minkowski vacuum $T_0 = 1/2\pi$ yields the analog of the Cardy formula in 4d (\ref{cardyd}) with $n=2$. Thus, unlike in two dimensions,  there is no simple, universal Cardy formula where the pressure depends only on the anomaly and therefore is independent of the coupling. In fact, we know that in general the pressure of a 4d CFT should depend on the coupling. For example, in the $N=4$ SYM at large $N_c$, the pressure in the free field theory limit differs from the strong coupling value determined via the AdS/CFT correspondence by the famous factor of $4/3$, while the
trace anomaly coefficients do not change from weak to strong coupling \cite{Henningson:1998gx}. Also, field theory calculations at weak coupling show that the pressure is indeed coupling dependent \cite{Gubser:1998nz}.

\subsection{Higher dimensions}

It is straightforward to generalize the above analysis to arbitrary even dimension.
First, since the anomalous action for the Weyl invariant terms $S_{\rm Weyl}$ (\ref{Weylinv}) contains only terms of at least second
power
in the curvatures of the background fields, it follows that
they do not contribute the stress-energy tensor in a flat background. This leaves only the contribution
from $S_{\rm Euler}$.
This action can be calculated from (\ref{anomaction}). For instance, in six dimensions it can be read from the dilaton action (B.17) in \cite{Elvang:2012st} by replacing
$\tau\to-\ln(T/T_0)$.
Evaluating it on a Rindler background gives
\begin{align}
P_{\rm Euler} = {1\over d-1}\rho_{\rm Euler} = -\frac{n a}{T_0^d} T^d \ ,
\end{align}
which is the value of the dilaton Lagrangian density
evaluated on Rindler space.

The zeroth order contribution from the invariant part
\begin{align}
L^{(0)}_{inv} = \sqrt{-g} p_0 T^d \ ,
\end{align}
evaluated on Rindler space,
also yields
\begin{align}
T^{\mu \nu} = p_0 T^d (d u^\mu u^\nu + g^{\mu \nu}) \ ,
\end{align}
and $P_0 = p_0 T^d, \rho_0=(d-1)P_0$.
As in four dimensions we parametrize the contributions of
the higher derivative invariant terms in the effective action by $\xi(\lambda,T_0)$.
Then,
\begin{align}
P_{\rm tot} = \left(p_0 - \xi(\lambda,T_0) - \frac{n a}{T_0^d} \right) T^d \ .
\end{align}
Demanding that the stress-energy tensor vanishes in the Minkowski vacuum $T_0 = 1/2\pi$ yields (\ref{cardyd}). In six dimensions
we get $n=24$.

\section*{Acknowledgements}

Y.O. would like to thank A. Yarom for a valuable discussion.
We also thank Adam Schwimmer for comments on the first version of this paper
and for pointing out the similarity with the dilation action.
This work is supported in part by the ISF center of excellence.

\end{document}